\documentclass[fleqn,twoside]{article}
\usepackage[headings]{espcrc2}
\usepackage{xspace}

\readRCS
$Id: Rouby_Xav.tex,v 1.2 2004/02/24 11:22:11 spepping Exp $
\usepackage{graphicx}
\usepackage[figuresright]{rotating}

\newcommand{\LHC}{\textsc{lhc}\xspace}
\newcommand{\CMS}{\textsc{cms}\xspace}
\newcommand{\HECTOR}{\textsc{Hector}\xspace}
\newcommand{\lumi}{\ensuremath{\textrm{cm}^{-2}\textrm{s}^{-1}}\xspace}

\newcommand{\AmS}{{\protect\the\textfont2
  A\kern-.1667em\lower.5ex\hbox{M}\kern-.125emS}}

\title{Tagging photon interactions at the \textsc{lhc}}

\author{X. Rouby\address[MCSD]{Universit\'e catholique de Louvain, Center for Particle Physics and Phenomenology (CP3), Louvain-la-Neuve, Belgium}%
\thanks{E-mail: xavier.rouby@uclouvain.be.}}

\runtitle{Proceedings of workshop on ``High energy photon collisions at the \LHC''}
\runauthor{X. Rouby}

\begin{document}

\begin{abstract}
Photon interactions at the \LHC result in striking final states with much lower
hadronic activity in the central detectors than for $pp$ interactions. In addition,
the elastic exchange of a photon leads to a proton scattered at almost zero-degree
angle. Tagging photon interactions relies on either the use of large rapidity gaps 
or on the detection of the scattered proton using very forward detectors. The
studies related to such detectors are presented, including their characterization,
their acceptance and reconstruction performance. Limitations due to the LHC beamline
misalignment and possible solutions are also given.
\vspace{1pc}
\end{abstract}

\maketitle

\section{Introduction}

A significant fraction of $pp$ collisions at the \textsc{lhc} will involve quasi-real photon 
interactions occurring at energies well beyond the electroweak energy scale~\cite{bib:piotr}. 
The \textsc{lhc} can therefore be considered to some extend as a high-energy photon-proton or 
photon-photon collider. The initial comprehensive studies of high energy photon interactions 
at the \LHC were recently reported~\cite{bib:nous,bib:severine,bib:nicolas,bib:tomek,bib:jerome,bib:jonathan,bib:these}. Clear identification of the remarkable photon-photon and photon-proton collisions 
out of the large sample of $pp$ events requires good experimental \textit{tagging} 
techniques. Two signatures characterizes the photon interactions: the presence of a very 
forward scattered proton and of a large pseudorapidity region of the detector, devoid of 
any hadronic activity, usually called \textit{large rapidity gap} (\textsc{lrg}), in the 
forward directions.

During the phase of low luminosity (i.e. significantly lower than $10^{33}~\lumi$),
the \textit{event pile-up} is negligible. Thanks to the colour flow in $pp$ interactions 
between the proton remnant and the hard hadronic final states, a simple way to suppress 
generic $pp$ interactions is to require \textsc{lrg}s. The \textsc{lrg} 
condition can be applied using a cut based on the energy measured in the forward detector 
containing the minimum forward activity ($3<|\eta|<5$). More details are given 
in~\cite{bib:severine}. Tagging photon interactions with \textsc{lrg} could rely only 
the central detector. But as the forward proton is not measured, the 
event final state is less constrained. At higher luminosity, the \textsc{lrg} technique 
cannot be used because of large event pile-up. Therefore the use of dedicated 
\textit{very forward detectors} (\textsc{vfd}s) is mandatory in order to retain $pp$ backgrounds low. 

Both \CMS and \textsc{atlas} experiments have extended their coverage in pseudorapidities
to the forward regions~\cite{bib:monika}. This includes near-beam calorimeters and tracking 
detectors. The transport of particles through the \LHC beamlines from the interaction point 
(\textsc{ip}) to the forward detectors is simulated with \textsc{Hector}~\cite{bib:hector}. 
The simulator is based on a linear approach of the beamline optics, implementing transport 
matrices from the optical element magnetic effective length, and with correction factors on 
magnetic strength for particles with non nominal energy. \textsc{Hector} deals with the 
computation of the position and angle of beam particles, and the limiting aperture of the 
optical elements. The measurement of the position and the angle of particles in dedicated 
near-beam very forward detectors, hundreds of meters away from the \textsc{ip}, helps in 
reconstructing the kinematics of the event at the related central detector. \HECTOR links 
the information from very forward detectors (\textsc{vfd}s) to the one from the central 
detector, by precise calculation of the particle trajectory. The measurement in the central
detector of exclusive dimuon final states, from $\gamma \gamma$ collisions or diffractive 
photoproduction of $\Upsilon$ mesons, can be used for a very precise calibration of the
\textsc{vfd}s~\cite{bib:these}.

\section{Simulation techniques}

\subsection{Physical description}

The simulator relies on a linear approach to single particle propagation. 
A beamline consists of a set of optical elements, amongst dipoles, quadrupoles, drifts, collimators, 
kickers and \textsc{vfd}s. The optical elements are described by their magnetic field, their length and their 
aperture. In turn, a set of particles, with all possible smearings of initial positions, angles or 
energies, is propagated by \textsc{Hector} through the beamline, particle by particle. The first terms 
of the Taylor expansion of the magnetic field can be interpreted as dipolar 
($\frac{1}{R} = \frac{e}{p} B_y$), quadrupolar ($k = \frac{e}{p} \frac{\partial B_y}{\partial x} $) 
and sextupolar fields. 
% The Taylor expansion of the vertical component of magnetic field $B_y$ around its central value is given by:
% $$\frac{e}{p} B_y(x) = \frac{e}{p} B_y + \frac{e}{p} \frac{\partial B_y}{\partial x} x + \frac{1}{2} \frac{e}{p} \frac{\partial^2 B_y}{\partial x^2} x^2 + \ldots$$
% where $p$ is the momentum of the particle and $e$ its electric charge. The terms of this sum are 
% interpreted as respectively dipolar ($\frac{1}{R} = \frac{e}{p} B_y$), quadrupolar 
% ($k = \frac{e}{p} \frac{\partial B_y}{\partial x} $) and sextupolar fields. 

In the co-moving coordinate 
system, neglecting small deviations ($x \ll R$, $y  \ll R$) and small momentum loss ($\Delta p \ll p$), 
this leads to the following equation of motions for a particle traveling along path length $s$ through 
a magnetic element \cite{KlausWille}:

\begin{equation} 
\label{eqmotion}
\Bigg\{ \begin{array}{l} x''(s) + \left( \frac{1}{R^2(s)} - k(s) \right) x(s) = \frac{1}{R(s)} \frac{\Delta p}{p} \\
	y''(s) + k(s) y(s) = 0.
\end{array}
\end{equation}

The solution $(x(s), x'(s), y(s), y'(s))$ to these equations can be expressed as a linear combination 
of the initial values $(x_0, x'_0, y_0, y'_0)$, where the rotation matrices are defined by the 
properties of the optical element (length and magnetic field). 
Each beam particle is represented by a phase space vector and each optical element by a transfer matrix by which the vector is multiplied. The propagation of a single particle is thus the \textit{rotation} of the phase space vector by the $n$ optical element matrices.

$$ X(s) = X(0) \underbrace{M_1 M_2 ... M_n}_{M_{\mathrm{beamline}}} $$

The whole beamline is modeled as a single transport matrix acting on each particle phase space vector 
(no intrabeam interactions). The optical element description also refers to its physical aperture. 
When a particle is propagated through an optical element, the compatibility between its trajectory
and the optical element aperture is checked.

\subsection{Implementation}
\label{Ecorrection}

Beam particles are described by a 6-components phase space vector $X = (x,x',y,y',E,1)$,
where 
$(x,x')$ and $(y,y')$ are the horizontal and vertical coordinates and angles; 
$E$ is the particle energy. The sixth component is just a factor used to 
add an angular kick on the particle momentum direction. 
The optical elements (dipoles, quadrupoles, drifts, $\ldots$) are modelled by $6 \times 6$ transport matrices:
$$
      \mathbf{M_{units}} =
      \left(
      \begin{array}{cccccc}
	\mathrm{A} & \mathrm{A} & 0 & 0 & 0 & 0 \\
	\mathrm{A} & \mathrm{A} & 0 & 0 & 0 & 0 \\
	0 & 0 & \mathrm{B} & \mathrm{B} & 0 & 0 \\
	0 & 0 & \mathrm{B} & \mathrm{B} & 0 & 0 \\
	\mathrm{D} & \mathrm{D} & 0 & 0 & 1 & 0 \\
	K & K & K & K & 0 & 1 \\
      \end{array}
      \right)
$$ where
\begin{itemize}
\item $\mathrm{A}$ (and $\mathrm{B}$) blocks refer to the action (focusing, defocusing, drift) on 
horizontal (and vertical, resp.) coordinate and angle.
\item $\mathrm{D}$ terms reflect the dispersion effects of the horizontal dipoles on off-momentum 
particles.
\item $K$ factors are the angular action of kickers.
\end{itemize}

The \textit{chromaticity}, or the energy dependence of the transport matrix, is implemented 
by rescaling the magnetic field terms ($R$, $k$, $K$) with a factor ($\frac{p}{p - \Delta p }$). 
The propagation of particles different from protons is also possible by rescaling these magnetic 
field terms:
\begin{equation} 
	k_i(\Delta E, q_p) = k_i \frac{p_0}{p_0 - \Delta p } 	
 	\frac{q_p}{q_\mathrm{proton}}, ~ k_i = R,k,K;
\end{equation}
where $q_p$ is the particle charge. Forward particles from the final state can be 
then propagated through the beamline via \textsc{Hector}. 

\section{Beamline simulation}

\begin{figure}[!h]
\centering
\includegraphics[width=1.13\columnwidth]{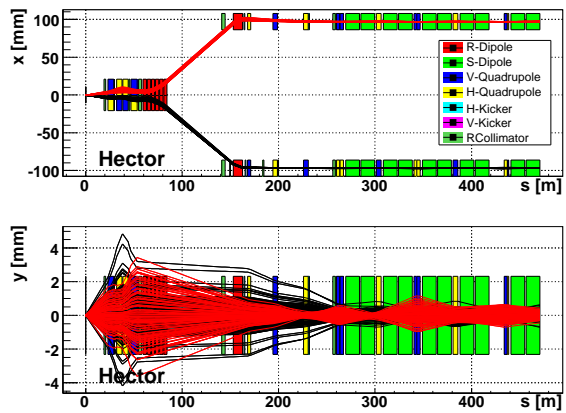}
\caption{Top and side views of both \LHC beams around the \textsc{ip}5. The \textsc{ip} is located at $s=0$ m. 
Beam 2 (red) flows from the right to the left and is seen before its passage at the \textsc{ip}. Beam 1 
(black) is seen downstream, passing from the left to the right after the crossing. In this graph, the 
bending effect of the sector magnets (S-Dipoles) has been switched off, thus rectifying the beam path 
to a straight line after $250 ~ \mathrm{m}$~\cite{bib:hector}.}
\label{2beams}
\end{figure}

Knowing the optics tables for both \LHC beams, their trajectories can be compared 
simultaneously, in both top and side views, for the two \LHC beams aside (incoming 
beam 2 and outgoing beam 1, at the \textsc{ip}5 and \textsc{ip}1: Fig. \ref{2beams}). 
The top view shows the beams on the horizontal plane, clearly depicting the crossing 
angle at the \textsc{ip}5, and the beam separation after $70 ~ \mathrm{m}$ away from the 
interaction point. The bending of the sector dipoles has been switched off in order 
to makes graphics more clear -- this is why both beams are straight and parallel after 
$250 ~ \mathrm{m}$. However, the optical elements have been shifted (without tilt) in 
the horizontal plane by the half of the beam separation distance, from $180 ~ \mathrm{m}$ 
away from the \textsc{ip}, in order to match the ideal beam path: a proton with nominal 
energy and on the ideal orbit should travel through optical elements in their geometrical 
center. The side view in turn shows the beams in the vertical plane. In addition, the 
major optical elements have been drawn: rectangular dipoles in red, sector dipoles 
in light green, and quadrupoles in yellow and blue.

The simulation of proton beams is needed for the validation of the simulator results,
by a direct comparison to the results of other simulators.

\subsection{Calibration}
The calibration of the reconstructed variables at forward detectors can be well 
maintained using the physics processes and the central detectors. 
At \textsc{hera}, for example, the elastic $\rho$ meson photoproduction was used where the momentum
of the scattered proton could be deduced from two decay charged pions using the central tracking. At
the \LHC, the two-photon exclusive production of dimuon pairs and the diffractive $\Upsilon$ 
photoproduction seems good calibration processes~\cite{bib:jonathan,bib:these}, with large statistics 
for the detectors at $420 ~ \textrm{m}$. The visible cross section of exclusive dimuons is large 
($\sigma \leq 7 ~ \textrm{pb}$), including the acceptance of central detectors.
The measurement of the final state muon pair provides an estimate of the proton energy loss $x=E_\gamma/E_\textrm{beam}$, which
can be matched with the proton position in the \textsc{vfd}s. From the dimuon longitudinal momentum 
$P_z$ and invariant mass $M_{\mu \mu}$, one gets
$$E_\gamma = \pm \frac{P_z}{2} + \frac{ \sqrt{(M_{\mu \mu})^2 + (P_z)^2 } }{2} .$$
This should allow for a run-by-run calibration of the scattered proton energy scale within 
a full acceptance range. Finally, the expected reconstruction power of central detectors is 
excellent for such dimuon events, giving the proton energy uncertainty about $10^{-4}$ per 
event \cite{bib:opus}. Recent results with the full similation of the detector show even
better resolutions, at the order $5 \times 10^{-6}$ (depending on $x$)~\cite{bib:these}. One should note, however, that using this process 
it is not possible to check the angular reconstruction, and that it has much more limited 
statistics within acceptance of the detectors at $220$~m.

\section{Very Forward Detectors}

In the following, the use of very forward detectors (\textsc{vfd}s) located at $220$~m and $420$~m from \textsc{ip5}
are discussed, as taggers for photon interactions.
The assumed location of the first detector is 
$(s=220-224\textrm{ m,} ~ x=2000 ~ \mu\textrm{m})$, and of the second one at 
$(s=420-428\textrm{ m,} ~ x=4000 ~ \mu\textrm{m})$. No hypothesis is made on their detection 
efficiency or their resolution, unless quoted. We consider here \textsc{vfd}s providing 2D-measurement 
($x$ and $y$ coordinates), each consisting in fact of two stations separated by $4$ and $8 ~\textrm{m}$ 
as a lever arm for the angle measurement, with no magnetic element in between. 

\subsection{Acceptance}
\begin{figure}[!h] %2D-acceptances
\centering
\includegraphics[width=\columnwidth]{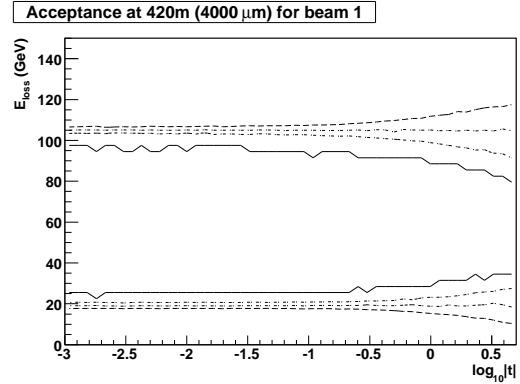}  
\caption{\textsc{vfd} proton acceptance for the \LHC beam 1 around the \textsc{ip}5. 
The \textsc{vfd} is located at ($s=420$~m, $x=4000 ~ \mu$m). This map shows contours of $25\%$, $50\%$, $75\%$ 
and (plain curve) $100\%$ acceptance. The acceptance is roughly rectangular, i.e. independent 
of $t$. The angular kick coming from the large momentum transfers leads to an 
increasing smearing of the graph lower edge~\cite{bib:hector}.} \label{RPacceptance2D}
\end{figure}

\begin{figure}[!h] %1D-acceptances
\centering
\includegraphics[width=\columnwidth]{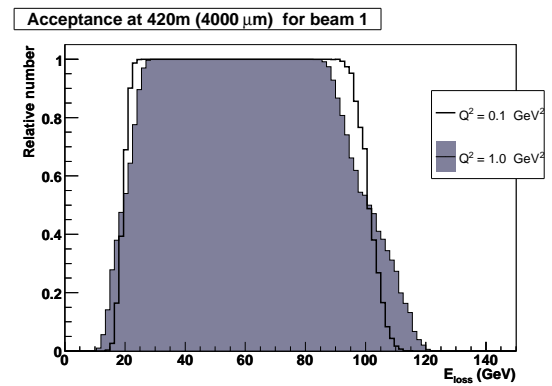} 
\caption[Roman pot 1D-acceptances ($E$)]{\textsc{vfd} acceptance as a function of energy loss, for two fixed 
virtualities (Reminder: $Q^2=-t$). See previous figure for more details~\cite{bib:hector}.}
\label{RPacceptance1D}
\end{figure}

Using \textsc{Hector}'s aperture description, it is possible to identify the 
characteristics of the protons that will hit the \textsc{vfd}. The exchange of a photon, 
leaving the proton intact, results in a proton energy loss ($E_{loss}$) and a scattering 
angle, directly linked to the four-momentum transfer squared ($t$), or equivalently to the 
photon virtuality ($Q^2$). The acceptance windows of the \textsc{vfd}s can be computed by 
performing scans in $(E_{loss},t)$ and computing the probabilities of reaching the detectors. 
The figure~\ref{RPacceptance2D} shows the contour plots of the detectors acceptance, in this 
$(E_{loss},t)$ plane. The \textsc{vfd} acceptances mostly depend on 
$E_{loss}$, and have a very small sensitivity in $t$, within a large $t$ range. 
Corresponding profile at fixed virtuality is shown in figure \ref{RPacceptance1D}.

The total diffractive cross-section at the \LHC is very large, resulting in a high rate of 
diffractive protons hitting the \textsc{vfd}s. As a result, it causes extremely high irradiation 
levels. About $10^{14}$ hits/cm$^2$ are expected per year at low luminosity. This illustrates the need
for very radiation hard detectors.

\subsection{Chromaticity grids}

\begin{figure}[!h] % Chromaticity grid X
\centering
\includegraphics[width=1.11\columnwidth]{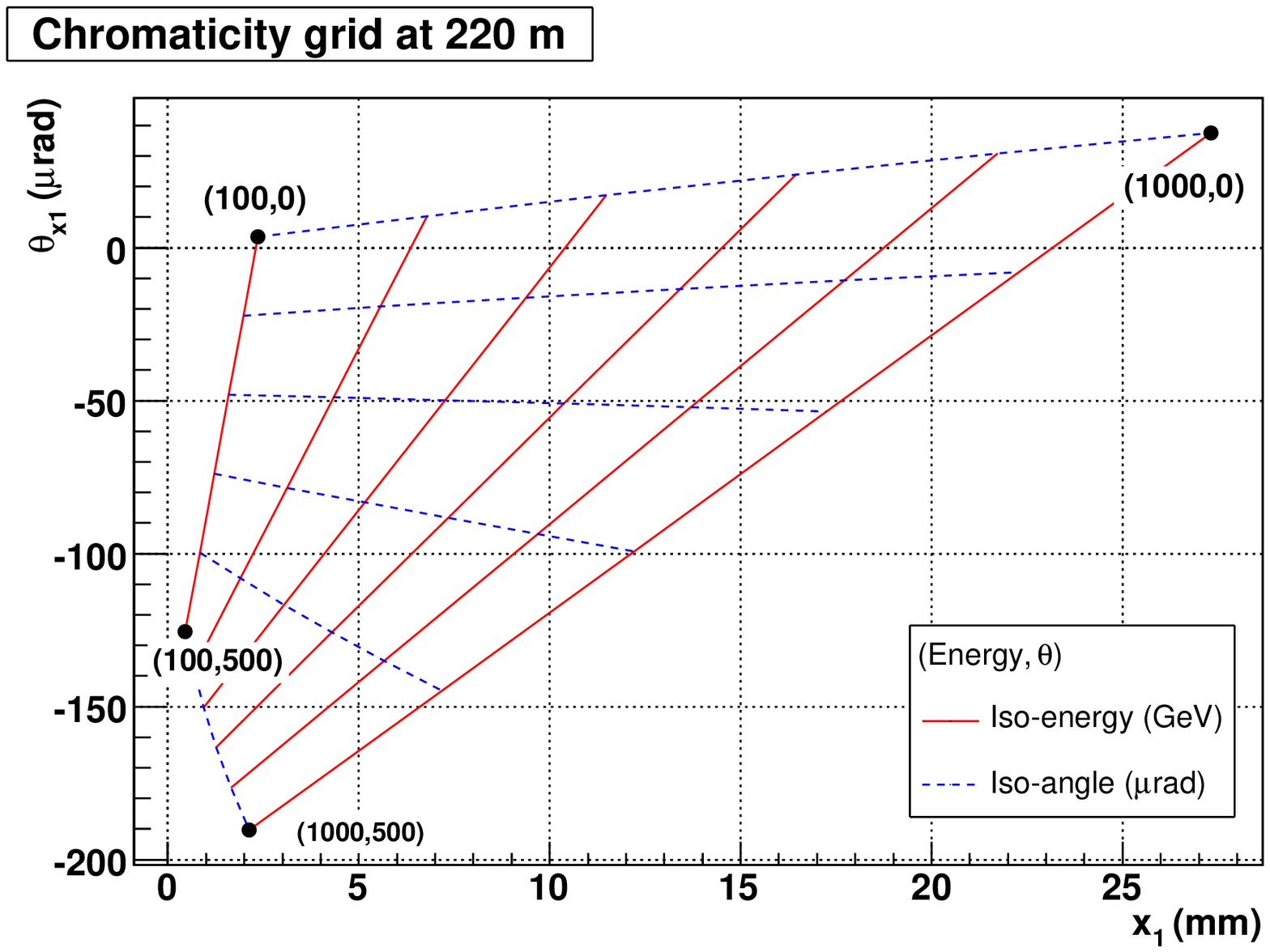} \\
\includegraphics[width=1.11\columnwidth]{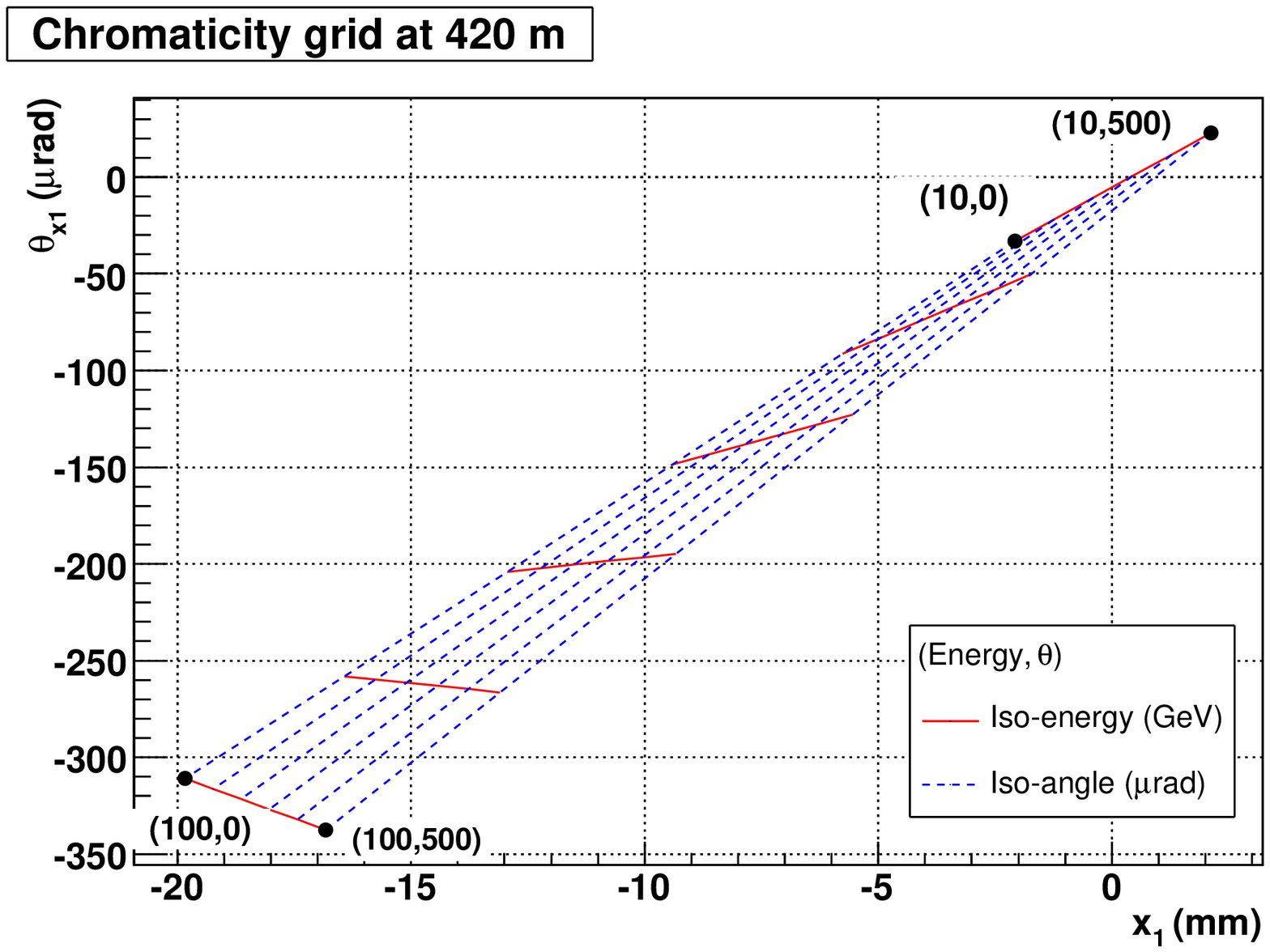} 
\caption[Chromaticity grids: Dependence on energy and virtuality loss at the \textsc{ip}]
{Chromaticity grids: iso-energy and iso-angle lines for the \textsc{vfd}s at $220$~m (\emph{above}) and $420$~m (\emph{below}) away from the \textsc{ip}5, for the \LHC beam 1. The graphs show the positions $x_1$ and angle $\theta_{x1}$ 
of protons, given the energy loss $[0; 1000]$~GeV and the angular kick of $[0; 500] ~ \mu$rad. 
The energy dependence of the transfer matrices induces a deformation of the grid, 
worsening the reconstruction power at higher angles~\cite{bib:hector}.}
\label{Chromaticityplotsx}
\end{figure}

Once the acceptance windows of very forward tracking detectors are defined, it is 
interesting to see matching between the proton variables at the \textsc{ip} and 
those measured by \textsc{vfd}s. Depending on their energy and angle at the \textsc{ip}, 
forward protons will hit the \textsc{vfd}s at various positions. Drawing iso-energy and 
iso-angle curves for a set of sample protons produces a grid in the measurement related 
variables, $(x_1 , x_2)$ or $(x_1, \theta)$. Due to optics of the \LHC beamlines, the 
grid unfolds itself in a much clearer way in the latter plane, and is almost invisible 
in the former one. The energy dependence of the transfer matrices implies a deformation 
of the grid -- without such a dependence, the grid would be a parallelogram. One should 
note, that uncertainty of the transverse position of the proton vertex at the \textsc{ip}
results effectively in smearing the chromaticity grids. Anyway, these chromaticity 
grids provide a straightforward tool for unfolding the energy and angle at the \textsc{ip} 
of the measured particle. The grid in figure~\ref{Chromaticityplotsx} is calculated in 
the energy range accessible to the \textsc{vfd}s.

\subsection{Reconstruction}

The reconstruction of the event kinematics is possible with \HECTOR, from the
\textsc{vfd} measurements. If a beam particle has exchanged a photon at the \textsc{ip}, 
one could reconstruct photon's energy ($E$) and virtuality ($Q^2$). The particle energy 
at a given position in the beamline is obtained from the measured particle position and 
angle within the matrix formalism by solving these equations: 
$$ \bigg\{ \begin{array}{ll}
	x_s = a_s x_0 + b_s x'_0 + d_s E \\
	x'_s = \alpha_s x_0 + \beta_s x'_0 + \gamma_s E\\
  \end{array} 
$$

The transfer matrix of the beamline yields the coefficients $a$, $b$, etc. The introduction of an 
energy dependence on the strength of optical elements refines the transfer matrix, becoming a 
function of $E$: $a_s(E)$, $b_s(E)$,... This dependence will introduce \emph{non linearities}. 

Neglecting the terms $a_s x_0$ and $b_s x'_0$ leads to significant sensitivity to the non-nominal 
values of the average vertex position and beam direction ($tilt$) at the \textsc{ip}. While the average 
vertex position can be very well measured using the central detectors, the beam tilt is more 
difficult to control -- it should be known to better than $10-20~\mu$rad to avoid causing a 
significant bias. One can reconstruct the energy and scattering angles using position measurements
at two detector stations at the same time. This requires to solve the equations for $x_s$ at both
detectors for E. The $a_s$, $b_s$ and $d_s$ coefficients of the transfer matrix depend on 
energy with rather complicated shapes, as they are products of many magnets matrices. One efficient 
way to get those is then simply to fit each coefficient as a function of energy. Various fitting 
functions were tried, but a quadratic fit proved to be sufficient to avoid any visible bias or 
resolution degradation.

Using these fitted coefficients, one can easily get a formula for $x_0$ as a function of $x$ at both 
detectors and of the energy. For each pair of detector $x$ coordinates, the used method is to numerically 
find the root ($x_0$ = 0) of the formula to get the energy corresponding to a $x_0$ = 0, and thus 
neglect the interaction point transverse extension.

This method allows to get reconstructed energy independently of the angle of the particle at the \textsc{ip}. 
Energy reconstruction resolutions does not degrade anymore if the particle transverse momentum $p_T$ rises, 
but the price to pay is that the detectors resolutions become critical. In particular, it means that the
uncertainty of the reconstructed angle at a given detector location should be better than the beam
angular divergences there (6 and 1.5 $\mu$rad, respectively). The transverse momentum can thus also be 
computed from the matrix coefficients once the energy has been reconstructed.

\begin{figure}[!h]
\centering
%\vspace{-1.5cm}
\includegraphics[width=0.9\columnwidth]{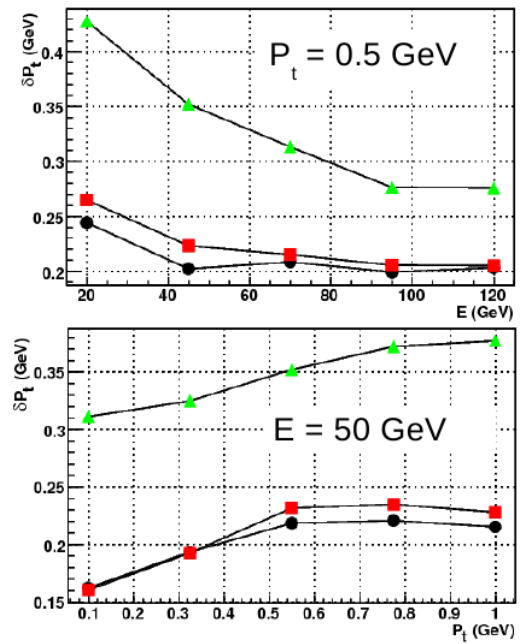}  
\caption[Reconstructed $p_T$ resolution using the advanced method]{Resolution of the 
	reconstruction of the particle transverse momentum $p_T$, as a function of the 
	energy loss (\emph{above}) and of the transverse momentum (\emph{below}), for 
	\textsc{vfd}s at $420$~m from the \textsc{ip}5. Dots 
	correspond to different scenarios of detectors resolutions, namely perfect
	detectors (\emph{circles}), $5~\mu$m (\emph{squares}) and $30~\mu$m (\emph{triangles}) spatial 
	resolution~\cite{bib:hector}.}
\label{amrecpt}
\end{figure}

\begin{figure}[!h]
\includegraphics[width=0.90\columnwidth]{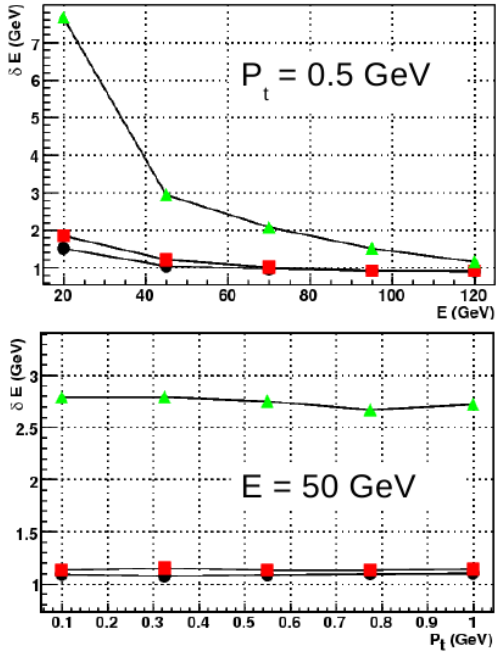}
\caption[Reconstructed Energy resolution using the advanced method]{Resolution 
    of the reconstruction of the particle energy loss $E$, as a function of the
    energy loss (\emph{above}) and of the transverse momentum (\emph{below}), for
    \textsc{vfd}s at $420$~m from the \textsc{ip}5.
    Dots
    correspond to different scenarios of detector resolutions, namely perfect
    detectors (\emph{circles}), $5~\mu$m (\emph{squares}) and $30~\mu$m (\emph{triangles}) spatial
    resolution~\cite{bib:hector}.}
\label{amrece}
\end{figure}

The reconstructed energy 
resolution stays very good even with non-negligible initial particle $p_T$. Figures 
\ref{amrecpt} and \ref{amrece} show the $p_T$ and energy dependence of resolutions on the 
transverse momentum and energy, for various detectors resolutions. As expected, the
energy resolution is independent of $p_T$ but is sometimes sensitive to the energy as 
expected from the chromaticity grids of Fig. \ref{Chromaticityplotsx}. The effect of the
detector resolution, which was absent for the simple reconstruction method described previously,
is now clearly visible even for an excellent resolution of 5~$\mu$m, especially for detectors 
at $220$~m from \textsc{ip}.

% end of advanced method section

In summary, the proton angular distribution at the \textsc{ip} affects the most the energy 
reconstruction. In contrast, the vertex lateral distribution has negligible
impact. As a result, the most important beam parameter, which can change 
run-to-run, is the beam tilt at the \textsc{ip}. In principle, it can be indirectly controlled
by the beam position monitors (\textsc{bpm}s) at $220$ and $420$~m, but more direct tilt measurements
are favored, as by using \textsc{bpm}s next to the \textsc{ip}, or by monitoring the direction of
neutral particle (photons, neutrons, or neutral pions) production in the zero-degree 
calorimeters (\textsc{zdc}s).

The energy resolution squared of the scattered proton $\sigma_E$ can be then
approximately decomposed into four terms:
$$ \sigma_E^2 = \sigma_0^2 + \sigma_{vtx}^2 + \sigma_{ang}^2 +  \sigma_{det}^2 , $$
where the nominal beam energy dispersion $\sigma_0\approx0.8$~GeV, the contribution due to the vertex spread 
$\sigma_{vtx}\approx0.7$~GeV at $420$~m and $1-2$~GeV at $220$~m, the contribution due to the detector resolution, neglecting the angular effects, $\sigma_{det}^2$ is small for resolutions better than $50~\mu$m, and the contribution due to the proton non-zero angle at the \textsc{ip}, $\sigma_{ang}$, which is very sensitive the angular reconstruction at the \textsc{vfd}s; 
even for a zero-degree scattering if one neglects effects due to the beam angular divergence at the \textsc{ip},
$\sigma_{ang}\approx1$~GeV at $420$~m and $\approx3$~GeV at $220$~m.

\section{Misalignments}
The misalignment of \LHC optical elements could have a significant impact on the measurements with very
forward detectors. As the deflection of the particle paths depends on their positions in quadrupoles, 
a misplacement of these optical elements implies a change in the nominal beam position.
In turn, as the accurate position measurement with the forward tracking detectors (as well as the 
information inferred from the segmentation of forward calorimeters) is referred to the ideal beam 
location, changing this reference results in a biased reconstruction of the measured particles.

\begin{figure}[!h]
\centering
\begin{tabular}{cc}
\includegraphics[width=\columnwidth]{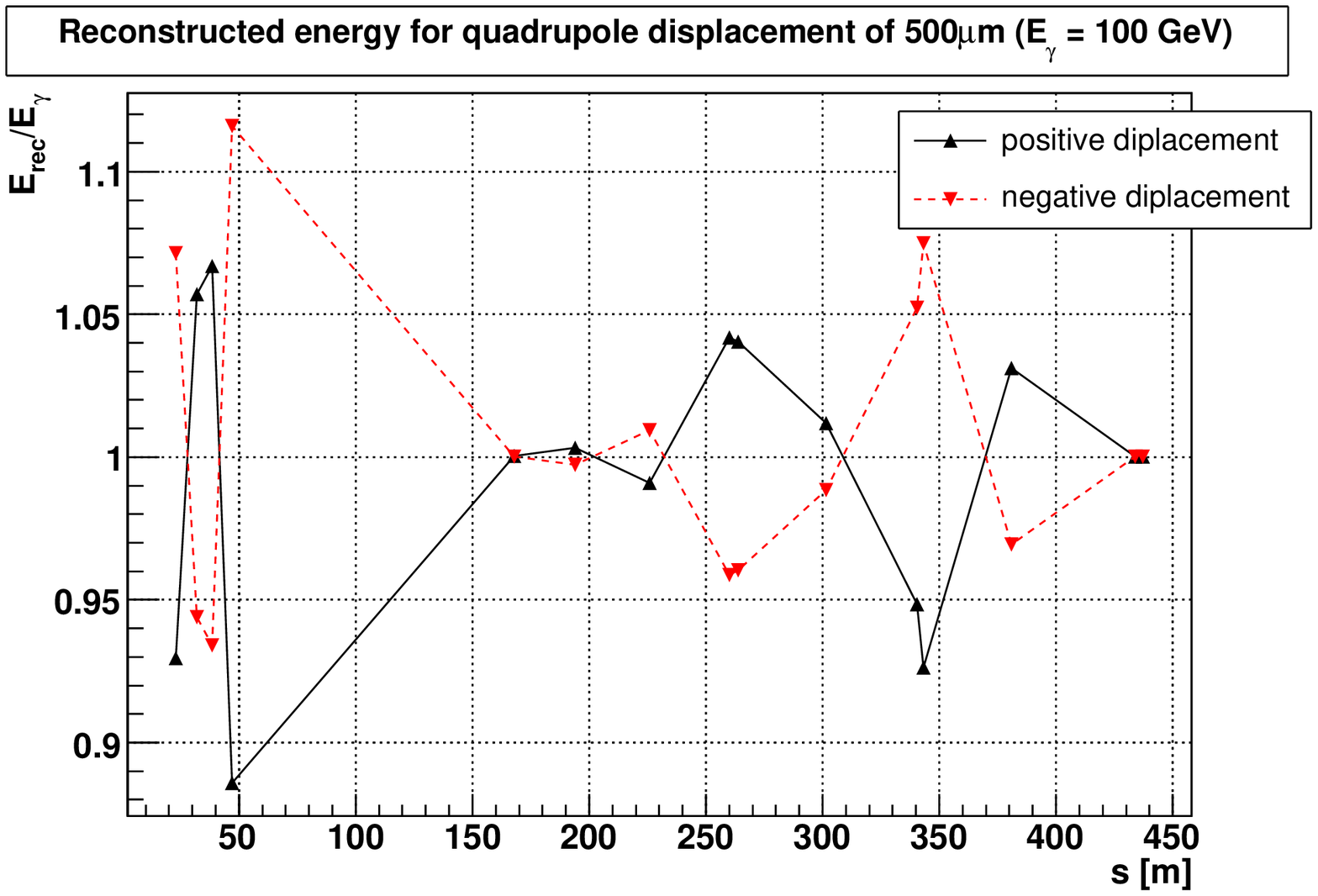} \\
\includegraphics[width=\columnwidth]{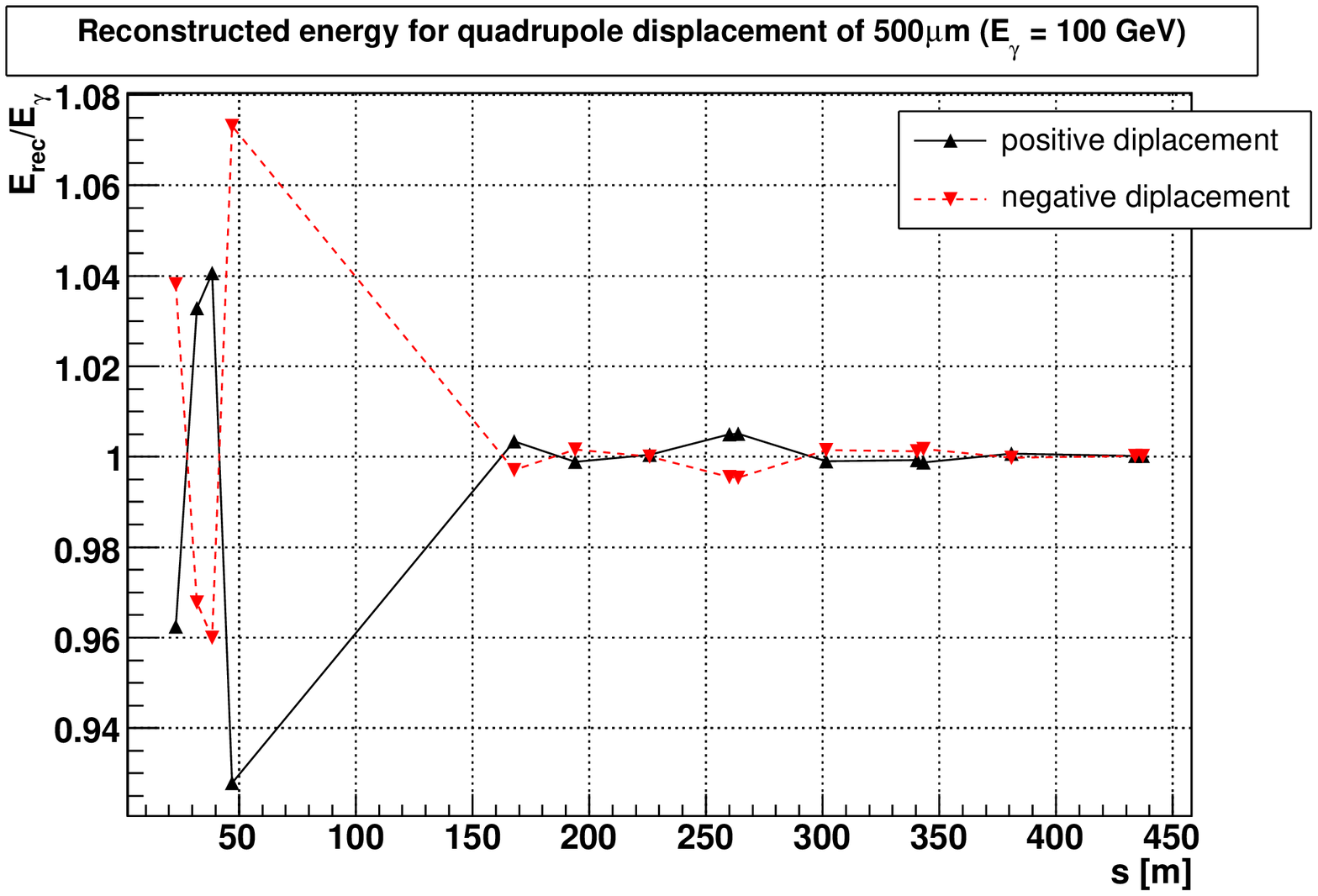} 
\end{tabular}
\caption[Reconstruction error due to the misalignment of quadrupoles]{The misalignment of the \LHC 
quadrupoles bias the energy reconstruction. 
%This effect can be partially corrected with the information of the exact beam position at a detector 
%location. 
The graphs show the bias for the reconstruction (with the trivial method) of a 100 GeV energy loss  
assuming the misaligned quadrupoles at various positions. Each element is separately shifted 
($500~\mu$m) , assuming a perfect alignment for the rest of the beamline. The impact of the 
misalignment can be important. Even a perfect knowledge of the actual beam position at the \textsc{vfd} 
(\emph{below}) does not compensate for this bias, depending on the position of the misplaced quadrupole~\cite{bib:hector}.}
\label{misalignment}
\end{figure}

Figure~\ref{misalignment} shows the impact of possible shifts ($0.5 ~\textrm{mm}$) of the 
beamline quadrupoles on the energy reconstruction with \textsc{vfd}s at 
$420 ~ \textrm{m}$. The reconstruction assumes ideal beamline in which only one quadrupole at a time 
is separately moved. Effects even higher than $10 ~ \%$ could be expected. 
Similar results are observed due to the beam tilts at the \textsc{ip}~\cite{bib:hector}.
One can partially correct for these effects using information from the beam position monitors, 
but better results are obtained using a physics calibration process like the two-photon muon pair 
exclusive production or the diffractive photoproduction of $\Upsilon$, at least in case of the 
\textsc{vfd}s at $420$~m~\cite{bib:these}.

\begin{figure}[!h]
\centering
\begin{tabular}{c}
\includegraphics[width=\columnwidth]{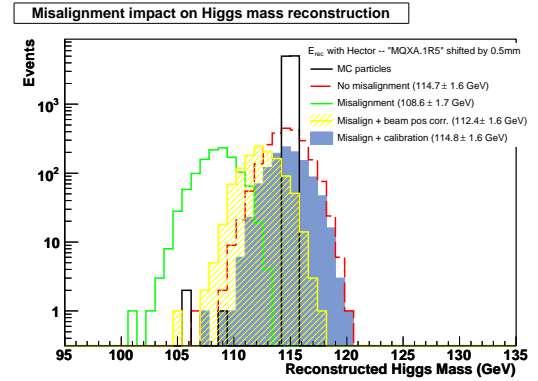}
\end{tabular}
\caption[Higgs two-photon exclusive production: error on reconstructed mass from misalignment]
% {Illustration of the effects in the energy reconstruction due to the misalignment of \LHC quadrupoles. 
% The graphs show the reconstructed Higgs boson mass in the two-photon exclusive production, using 
% energy of two forward scattered protons. In the upper plot, a quadrupole 
% (\textsc{mqm9r5}, $s= 347$ m) close to the detector has been shifted by $100~\mu \textrm{m}$. Misaligning an optical 
% element (\textsc{mqxa1r5}, $s=29$ m) close to the \textsc{ip} leads to a loss of acceptance (lower plot). 
% The reconstructed values including the correction due to the dimuon calibration 
% is also plotted. In brackets, the average reconstructed mass and its resolution are given, 
% without including the beam energy dispersion~\cite{bib:hector}.}
{Illustration of the effects in the energy reconstruction due to the misalignment of \LHC quadrupoles. 
The graphs show the reconstructed Higgs boson mass in the two-photon exclusive production, using 
energy of two forward scattered protons. Misaligning a quadrupole (\textsc{mqxa1r5}, $s=29$ m) close to the \textsc{ip} leads to a loss of acceptance. The reconstructed values including the correction due to the dimuon calibration are also plotted. In brackets, the average reconstructed mass and its resolution are given, 
without including the beam energy dispersion~\cite{bib:hector}.}
\label{misalignment_higgs}
\end{figure}

These misalignment effects and the corrections are illustrated (Fig. \ref{misalignment_higgs}) by 
the study of two-photon exclusive production of the SM Higgs boson 
($\textrm{pp} \rightarrow \textrm{pp}(\gamma \gamma) \textrm{H}$) with $\textrm{M}_{\textrm{H}} = 115 ~ \textrm{GeV}$.
The measurement of the energy of two scattered protons straightforwardly yields the boson mass, by means of the
so-called \textit{missing mass method}. 
As a consequence, an uncorrected measurement with misalignment leads to a bad mass calculation.
%The effect of the beam dispersion is a direct smearing. 
A quadrupole (\textsc{mqxa1r5}, $s=29$ m) close to the \textsc{ip} is shifted by  $500~\mu \textrm{m}$.
The misalignment-induced change in the \textsc{vfd} acceptance is visible. The limitations of the beam-position-based corrections are clearly visible, even assuming no systematic errors, 
while the muon-calibration stays unbiased (though only a relatively small sample of 700 dimuon events was used to
get the correction factors).

\section{Summary and outlook}

Photon interactions at the \LHC result in striking final states with much lower
hadronic activity in the central detectors than for $pp$ interactions. While 
forward rapidity gaps can be used for their tagging in a low pile-up environment,
the use of very forward detectors is mandatory for luminosity starting at 
$\mathcal{L}= 2\times 10^{33}~\textrm{cm}^{-2}\textrm{s}^{-1}$. The good simulation of 
particle transport in the \LHC beamline is very important. It allows to
characterize the detectors and to reconstruct the event from the scattered proton 
measurement. Impact of beamline misalignement has been studied as well as the calibration 
possibilities from the observation of exclusive dimuon final states.

\end{document}